\documentclass[prc,aps,superscriptaddress,nofootinbib,showpacs,
onecolumn]{revtex4}
\usepackage{graphicx}
\usepackage{amsfonts}
\usepackage{amssymb}
\usepackage{amsmath}
\usepackage{natbib}
\usepackage{dcolumn}
\usepackage{bm} 
\newcommand{\bwt}{\begin{widetext}}
\newcommand{\ewt}{\end{widetext}}
\newcommand{\beq}{\begin{equation}}
\newcommand{\eeq}{\end{equation}}
\newcommand{\bea}{\begin{eqnarray}}
\newcommand{\eea}{\end{eqnarray}}
\begin{document}
\title{Warm and dense stellar matter under strong magnetic fields}
\author{A. Rabhi}
\affiliation{Centro de F\' {\i}sica Computacional, Department of Physics, University of Coimbra, 3004-516 Coimbra, Portugal} 
\affiliation{Laboratoire de Physique de la Mati\'ere Condens\'ee, Facult\'e
des Sciences de Tunis, Campus Universitaire, Le Belv\'ed\'ere-1060, Tunisia}
\author{P. K. Panda}
\affiliation{Department of Physics, C.V. Raman College of Engneering,
Vidya Nagar, Bhubaneswar-752054, India}
\affiliation{Centro de F\' {\i}sica Computacional, Department of Physics, University of Coimbra, 3004-516 Coimbra, Portugal} 
\author{C. Provid\^encia}
\affiliation{Centro de F\' {\i}sica Computacional, Department of Physics, University of Coimbra, 3004-516 Coimbra, Portugal} 
\date{\today}    
\begin{abstract}
We investigate the effects of strong magnetic fields on the equation of
state of warm stellar matter as it may occur in a protoneutron star. Both neutrino free and neutrino trapped matter at a fixed entropy per baryon are analyzed. A relativistic meanfield nuclear model, including the possibility of hyperon formation, is considered. A density dependent magnetic field with the magnitude
$10^{15}$ G at the surface and not more than $3\times 10^{18}$ G at the center is considered. The magnetic field gives rise to a neutrino suppression, mainly at low
densities, in matter with trapped neutrinos. It is shown that an hybrid protoneutron star will not evolve to a low mass blackhole if the magnetic field is strong enough and the magnetic field does not decay. However,  the decay of the magnetic field after cooling may give rise to the formation of a low mass blackhole. 
\end{abstract}
\maketitle

\section{Introduction}
Neutron stars with very strong magnetic fields are known as magnetars 
\cite{duncan,usov,pacz}. Recent observations suggest that anomalous 
x-ray pulsars (AXPs) and soft $\gamma$-ray repeaters (SGRs) are candidates 
for magnetars \cite{kouv,hurley,mareghetti}. The magnetic
field at the surface of the magnetars may be as strong as $10^{14}$-
$10^{15}$ G and magnetars are warm, young stars, $\sim~ 1$ kyear old.
On the other hand, it is estimated that the interior field in neutron stars
may be as large as $10^{18}$ G \cite{shap83}.
Ferrario and Wickammasinghe \cite{ferrario} suggest that the extra strong 
magnetic field of the magnetars results from their stellar progenitor 
with a high magnetic field core. Iwazaki \cite{iwazaki} proposed that the huge 
magnetic field of the magnetars is some color ferromagnetism of quark matter. 
Recently, Vink and Kuiper \cite{vink} have suggested that the magnetars orginate 
from rapid rotating protoneutron stars.

Motivated by the existence of strong magnetic fields in neutron stars, 
theoretical studies on the effects of extremely large fields on dense matter
and neutron stars have been carried out by many authors 
\cite{chakrabarty,broderick,cardall,Mao03,aziz08}. For densities
below twice normal nuclear matter density $(\rho \sim~0.153$ fm$^{-3})$, the
matter consists  only of nucleons and leptons. However, for baryon densities
above $2\rho_0$, the equation of state (EOS) and the composition of matter
is much less certain and the strangeness degrees of freedom should be taken 
into account either through the inclusion of hyperons, kaon condensation or a
deconfinement phase transition into strange quark matter. The inclusion
of hyperons and kaon condensation 
\cite{broderick1,suh,yue,yue1,aziz09}, tends
to soften the EOS at high densities.

In \cite{broderick1}, it was shown that the threshold densities of hyperons 
can be significantly altered by strong magnetic fields. Similar conclusions
were obtained in \cite{yue} where the strangeness was included through an 
antikaon condensation or in \cite{yue1} where not only hyperons but also the 
strange mesons $\sigma^*$ and $\phi$ were included in the EOS.

The effects of the magnetic field on the structure and composition of a neutron
star allowing quark--hadron phase transition has been studied in \cite{pal}.
A strong magnetic field makes the overall EOS softer.
However, due to the positive contribution of the magnetic field pressure to 
the total EOS,   an increase is of the maximum mass is predicted \cite{broderick1,hpais,aziz09,njl}.

Protoneutron stars appear as the  outcome of the gravitational
collapse of a massive star. During its early evolution, the 
protoneutron star, with an entropy per baryon of the order of
1  (units of the Bolztmann constant), contains trapped neutrinos. This stage is followed by
a deleptonization period, during which the core is heated up and reaches an entropy per
particle  ${\mathfrak{s}}\sim 2$, before cooling occurs. During the cooling stage, exotic degrees of freedom
such as hyperons or a kaon condensate, will appear \cite{prakash97}.

In this paper, we focus on the properties of warm stellar matter under a strong 
magnetic field which is composed of a chemically equilibrated and charge
neutral mixture of nucleons, hyperons and leptons. 
We will consider both neutrino free matter and matter with trapped neutrinos.
The effect of the  magnetic field 
on the  composition of warm stellar matter, both with trapped neutrinos and neutrino free,
and the properties of the equation of state (EOS) will be discussed.
Both the  Landau quantization which affects the charged particles and  the incorporation of the nucleon anomalous 
magnetic moments (AMM)  for field strengths $B> 10^5 B_e^c$ ($B_e^c=4.414\times
10^{13}$ G is the electron critical field) have important effects. 

This work is organised as follows: In section II, we derive
the equation of state for hadronic matter at finite temperature with magnetic
field. We present the results in section III. Finally, the conclusions are 
summarized in section IV.

\section{Hadron matter equation of state}
For the description of the EOS of stellar matter, we employ a field-theoretical 
approach in which the baryons interact via the exchange of $\sigma-\omega-\rho$ 
mesons in the presence of a static magnetic field $B$ along the $z$-axis 
\cite{bb,gm91,glen00}. 

The Lagrangian density of the non-linear Walecka model (NLWM) we consider has the form \cite{gm91}  
\beq
{\cal L}= \sum_{b}{\cal L}_{b} + {\cal L}_{m}+ \sum_{l}{\cal L}_{l}.
\label{lan}
\eeq
The baryon, lepton ($l$=$e$, $\mu$), and meson ($\sigma$, $\omega$ and  $\rho$) 
Lagrangians are given by \cite{bb,glen00}
\bwt
\bea
{\cal L}_{b}&=&\bar{\Psi}_{b}\left(i\gamma_{\mu}\partial^{\mu}-
q_{b}\gamma_{\mu}A^{\mu}- m_{b}+g_{\sigma  b}\sigma
-g_{\omega  b}\gamma_{\mu}\omega^{\mu}-g_{\rho b}\tau_{3_{b}}
\gamma_{\mu}\rho^{\mu}-\frac{1}{2}\mu_{N}\kappa_{b}\sigma_{\mu \nu} 
F^{\mu \nu}\right )\Psi_{b} \nonumber\\
{\cal L}_{l}&=& \bar{\psi}_{l}\left(i\gamma_{\mu}\partial^{\mu}-q_{l}\gamma_{\mu}A^{\mu}
-m_{l}\right )\psi_{l} \nonumber\\
{\cal L}_{m}&=&\frac{1}{2}\partial_{\mu}\sigma \partial^{\mu}\sigma
-\frac{1}{2}m^{2}_{\sigma}\sigma^{2}-U\left(\sigma \right)
+\frac{1}{2}m^{2}_{\omega}\omega_{\mu}\omega^{\mu}
-\frac{1}{4}\Omega^{\mu \nu} \Omega_{\mu \nu}  \cr
&-&\frac{1}{4} F^{\mu \nu}F_{\mu \nu}
+\frac{1}{2}m^{2}_{\rho}\boldsymbol{\rho}_{\mu}\boldsymbol{\rho}^{\mu}
-\frac{1}{4}  \mathbf{R}^{\mu\nu}\mathbf{R}_{\mu\nu}
\label{lagran}
\eea
\ewt
where $\Psi_{b}$ and $\psi_{l}$ are the baryon and lepton Dirac fields, 
respectively. The index $b$ runs over the eight lightest baryons $n$, $p$, 
$\Lambda$, $\Sigma^-$, $\Sigma^0$, $\Sigma^+$,  $\Xi^-$ and $\Xi^0$,
and the sum on $l$ is over 
electrons and muons ($e^{-}$ and $\mu^{-}$). $\sigma$, $\omega$, and $\rho$ 
represent the scalar, vector, and vector-isovector meson fields, which describe the nuclear interaction and $A^\mu=(0,0,Bx,0)$ 
refers to a external magnetic field along the z-axis. The baryon mass and 
isospin projection are denoted by $m_{b}$ and $\tau_{3_b}$, respectively. 
The mesonic and electromagnetic field tensors are given by their usual 
expressions: 
$\Omega_{\mu \nu}=\partial_{\mu}\omega_{\nu}-\partial_{\nu}\omega_{\mu}$, 
$\boldsymbol{R}_{\mu \nu}=\partial_{\mu}\boldsymbol{\rho}_{\nu}-
\partial_{\nu}\boldsymbol{\rho}_{\mu}$, and  
$F_{\mu \nu}=\partial_{\mu}A_{\nu}-\partial_{\nu}A_{\mu}$. 
The baryon AMM are introduced via the coupling 
of the baryons to the electromagnetic field tensor with $\sigma_{\mu \nu}
=\frac{i}{2}\left[\gamma_{\mu}, \gamma_{\nu}\right] $ and strength 
$\kappa_{b}$. The electromagnetic field is assumed to be externally generated 
(and thus has no associated field equation), and only frozen-field 
configurations will be considered. The  interaction couplings are denoted by 
$g$, the electromagnetic couplings by $q$ and the baryons, mesons and leptons 
masses by $m$. The scalar self-interaction is taken to be of the form
\beq
U\left(\sigma \right)=\frac{1}{3}b m_n \left(g_{\sigma N}\sigma \right)^3+ 
\frac{1}{4}c \left(g_{\sigma N}\sigma \right)^4.
\eeq
From the Lagrangian density in Eq.~(\ref{lan}), we obtain the following meson field equations in the mean-field approximation 
\bea
m^{2}_{\sigma} \sigma  +\frac{\partial U\left(\sigma \right)}{\partial\sigma}
&=&\sum_{b}g_{\sigma b}\rho^{s}_{b}=g_{\sigma N}\sum_{b}x_{\sigma b}
\rho^{s}_{b} \label{mes1} \\
m^{2}_{\omega} \omega^{0} &=& \sum_{b}g_{\omega b}\rho^{v}_{b}=
g_{\omega N}\sum_{b}x_{\omega b}\rho^{v}_{b} \label{mes2} \\
m^{2}_{\rho} \rho^{0} &=&\sum_{b}g_{\rho b}\tau_{3_{b}}\rho^{v}_{b}=
g_{\rho N}\sum_{b}x_{\rho b}\tau_{3_{b}}\rho^{v}_{b} \label{mes3}
\eea
where $\sigma=\left\langle  \sigma \right\rangle,\; \omega^{0}=
\left\langle \omega^{0} \right\rangle\;\hbox{and}\;
\rho=  \left\langle\rho^{0} \right\rangle$ 
are the nonvanishing expectation values of the mesons fields in uniform matter, and $\rho^v_b$ and $\rho^s_b$ 
are, respectively, the baryon vector and scalar densities.
 
The Dirac equations for baryons and leptons are, respectively, given by 
\bwt
\bea
\big[i\gamma_{\mu}\partial^{\mu}-q_{b}\gamma_{\mu}A^{\mu}-m^{*}_{b}
-\gamma_{0}\left(g_{\omega}\omega^{0}
+g_{\rho}\tau_{3_{b}}\rho^{0}\right) 
-\frac{1}{2}\mu_{N}\kappa_{b}\sigma_{\mu \nu} F^{\mu \nu}\big] \Psi_{b}
&=&0 \label{MFbary}\\
\left(i\gamma_{\mu}\partial^{\mu}-q_l\gamma_{\mu}A^{\mu}-m_l \right) 
\psi_{l}&=&0 \label{MFlep}
\eea
\ewt
where the effective baryon masses are given by 
\beq
m^{*}_{b}=m_{b}-g_{\sigma}\sigma \label{effmass},
\eeq 
and $\rho^{s}_b$ and $\rho^{v}_b$ are the scalar density and the vector
density, respectively. For a stellar matter consisting of 
a $\beta$-equilibrium mixture of baryons and leptons, the following 
equilibrium conditions must be imposed:
\bea \mu_b=q_b \, \mu_n - q_l \, \mu_e,\label{free}\eea
\beq \mu_{\mu}=\mu_{e}, \label{beta}
\eeq
where $\mu_i$ is the chemical potential of species $i$. The electric charge 
neutrality condition is expressed by
\beq
\sum_{b} q_{b} \rho^{v}_{b}+\sum_{l}q_{l} \rho^{v}_{l}=0,
\label{neutra}
\eeq
where $ \rho^{v}_{i}$ is the number density of particle $i$.
If trapped neutrinos are included, we replace $\mu_{e}\rightarrow
\mu_{e}-\mu_{\nu_e}$ in the above equations,
\bea 
\mu_b&=&q_b \, \mu_n - q_l \, \left(\mu_e-\mu_{\nu_e}\right).\\
\mu_{\mu}-\mu_{\nu_\mu}&=&\mu_{e}-\mu_{\nu_e},\label{trap}
\eea
where $\mu_{\nu_e}$ is the electron neutrino chemical potential.
The introduction of additional variables, the neutrino chemical potentials,
requires additional constraints, which we supply by fixing the lepton fraction
$Y_{Le}=Y_{e}+Y_{\nu_{e}}=0.4$ \cite{prakash97,burrows86}. Since no
muons are  present before and during the supernova explosion, the constraint
$Y_{L\mu}=Y_{\mu}+Y_{\nu_{\mu}}=0$ must be imposed. However, because the muon fraction is very
small in matter with trapped neutrinos, we only include muons in neutrino free matter.

The energy spectra for charged baryons, neutral baryons  and leptons 
(electrons and muons) are, respectively, given by
\bea
E^{b}_{\nu, s}&=&\tilde \varepsilon^b_{\nu, s}+g_{\omega b} \omega^{0}
+\tau_{3_{b}}g_{\rho b}\rho^{0} \label{enspc1}\\
E^{b}_{s}&=&\tilde \varepsilon^b_{s}+g_{\omega b} \omega^{0}+
\tau_{3_{b}}g_{\rho b}\rho^{0}\label{enspc2} \\
E^{l}_{\nu, s}&=&\tilde \varepsilon^l_{\nu, s}= \sqrt{\left(k^{l}_{\parallel}\right)^{2}+m_{l}^{2}+
2\nu |q_{l}| B}\label{enspc3}
\eea
where,
\bea
\tilde\varepsilon^{b}_{\nu, s}&=&\sqrt{\left(k^{b}_{\parallel}\right)^{2}+
\left(\sqrt{m^{* 2}_{b}+2\nu |q_{b}|B}-s\mu_{N}\kappa_{b}B \right)^{2}} \\
\tilde \varepsilon^{b}_{s}&=& \sqrt{\left(k^{b}_{\parallel}\right)^{2} +
\left(\sqrt{m^{* 2}_{b}+\left(k^{b}_{\bot}\right)^{2} }-s\mu_{N}\kappa_{b}B 
\right)^{2}}
\eea
and $\nu=n+\frac{1}{2}-sgn(q)\frac{s}{2}=0, 1, 2, \ldots$ enumerates the 
Landau levels (LL) of the fermions with electric charge $q$, the quantum 
number $s$ is $+1$ for spin up and $-1$ for spin down cases and $k_\parallel, \, k_\bot$ are,
respectively, the momentum components parallel and perpendicular to the magnetic field.

At finite temperature the
occupation number distribution functions are given, respectively for charged and neutral baryons,  by
\bwt
\bea
f^b_{k,\nu, s} &=& \frac{1}{1+\exp\left[\beta(\tilde\varepsilon^b_{\nu, s} 
-\mu^{*}_{b}) \right]} \qquad
\bar{f}^b_{k,\nu, s} = \frac{1}{1+\exp\left[\beta(\tilde\varepsilon^b_{\nu, s} 
+\mu^{*}_{b} )\right]},\\
f^b_{k, s} &=& \frac{1}{1+\exp\left[\beta(\tilde\varepsilon^b_{s} 
-\mu^{*}_{b}) \right]} \qquad
\bar{f}^b_{k, s} = \frac{1}{1+\exp\left[\beta(\tilde\varepsilon^b_{s} 
+\mu^{*}_{b} )\right]},
\eea
and for the charged leptons
\bea
f^l_{k,\nu, s} &=& \frac{1}{1+\exp\left[\beta(\tilde\varepsilon^l_{\nu, s} 
-\mu_{l}) \right]} \qquad
\bar{f}^l_{k,\nu, s} = \frac{1}{1+\exp\left[\beta(\tilde\varepsilon^l_{\nu, s} 
+\mu_{l} )\right]},
\eea
\ewt
where the baryon effective chemical potential $(\mu_b)^*$ is given by 
\bea
\mu^{*}_{b}&=&\mu_{b}-g_{\omega b}\omega^{0}-g_{\rho b}\tau_{3_{b}}\rho^{0}.
\eea
For the charged baryons, the scalar and vector densities are, respectively, given by
\bea
\rho^{s}_{b}&=&\frac{|q_{b}|Bm^{*}_{b}}{2\pi^{2}}\sum_{\nu, s}
\int^{\infty}_{0}\frac{dk^b_{\parallel}}{\sqrt{(k^{b}_{\parallel})^2
+(\bar m^c_{b})^2}}\left( f^b_{k,\nu, s}+\bar{f}^b_{k, \nu, s}\right), \cr
\rho^{v}_{b}&=&\frac{|q_{b}|B}{2\pi^{2}}\sum_{\nu, s}\int^{\infty}_{0}
dk^b_{\parallel}\left( f^b_{k,\nu, s}-\bar{f}^b_{k,\nu, s}\right),
\eea
where
 we have introduced the effective mass 
\beq
\bar m^c_{b}=\sqrt{m^{* 2}_{b}+2\nu |q_{b}|B}-s\mu_{N}\kappa_{b}B.
\label{mc}
\eeq
the expressions of

For the neutral baryons, the scalar and vector densities of the neutral baryon $b$ are, respectively, given by
\bwt
\bea
\rho^{s}_{b}&=&\frac{1}{2\pi^{2}}\sum_{s}\int^{\infty}_{0}
k^{b}_{\bot}dk^{b}_{\bot}\left(1-\frac{s\mu_{N}\kappa_{b}B}
{\sqrt{m^{* 2}_{b}+\left(k^{b}_{\bot}\right)^{2}}} \right)  
\int^{\infty}_{0}\, dk^{b}_{\parallel}
\frac{m^*_b}{\tilde \varepsilon^{b}_{s}}\left(
f^b_{k, s}+\bar{f}^b_{k, s}\right),  \cr
\rho^{v}_{b}&=&\frac{1}{2\pi^{2}}\sum_{s}\int^{\infty}_{0}k^{b}_{\bot}
dk^{b}_{\bot} \int^{\infty}_{0}\, dk^{b}_{\parallel}
\left( f^b_{k, s}-\bar{f}^b_{k, s}\right).  
\eea
\ewt
The vector density of the charged leptons is given by
\beq
\rho^{v}_{l}=\frac{|q_{l}|B}{2\pi^{2}}\sum_{\nu, s}\int^{\infty}_{0}
dk^{l}_{\parallel}\left( f^l_{k,\nu, s}-\bar{f}^l_{k,\nu, s}\right),
\eeq
and for neutrinos  by 
\beq
\rho^{v}_{\nu_e} = \frac{1}{2 \pi^2}\int^{\infty}_{0}k^2 dk 
\left( f^\nu_{k, s}-\bar{f}^\nu_{k, s}\right).
\eeq

We solve the coupled Eqs.~(\ref{mes1})-(\ref{MFlep}) self-consistently at a 
given baryon density $\rho=\sum_{b}\rho^{v}_{b}$ in the presence of 
a strong magnetic field. The energy density of stellar matter is given by 
\beq
\varepsilon_m=\sum_{b} \varepsilon_{b}+\sum_{l=e,\mu}\varepsilon_{l}+\frac{1}
{2}m^{2}_{\sigma}\sigma^{2}+U\left(\sigma \right) +
\frac{1}{2}m^{2}_{\omega}\omega^{2}_{0}+\frac{1}{2}m^{2}_{\rho}\rho^{2}_{0}.
\label{ener}
\eeq
where the energy densities of charged baryons $\varepsilon_b^c$, neutral baryons
$\varepsilon_b^n$, and leptons  $\varepsilon_l$ have, respectively, the
following forms
\bwt
\bea
\varepsilon^c_{b}&=&\frac{|q_{b}|B}{2\pi^ {2}}\sum_{\nu, s}\int^{\infty}_{0}
dk^b_{\parallel}\sqrt{(k^{b}_{\parallel})^2+(\bar m^c_{b})^2}\left(
f^b_{k,\nu, s}+\bar{f}^b_{k,\nu, s}\right) ,\cr
\varepsilon^n_{b}&=&\frac{1}{2\pi^ {2}}\sum_{s}\int^{\infty}_{0}k^{b}_{\bot}
dk^{b}_{\bot} \int^{\infty}_{0} dk^{b}_{\parallel}  
\sqrt{\left(k^{b}_{\parallel}\right)^{2} +\left(\sqrt{m^{* 2}_{b}+
\left(k^{b}_{\bot}\right)^{2} }-s\mu_{N}\kappa_{b}B 
\right)^{2}} \left( f^b_{k, s}+\bar{f}^b_{k, s}\right) \cr
\varepsilon_{l}&=&\frac{|q_{l}|B}{2\pi^ {2}}\sum_{\nu, s}\int^{\infty}_{0}
dk^{l}_{\parallel}\sqrt{(k^{l}_{\parallel})^2+m_{l}^{2}+2\nu |q_{l}| B}
\left( f^l_{k, \nu, s}+\bar{f}^l_{k,\nu, s}\right ).
\eea
\ewt
The thermodynamical grand potential and the free energy density are defined as
\beq
\Omega= {\cal F}-\sum_{b}\mu_{b}\rho^{v}_{b},\quad\quad {\cal F}=\varepsilon_m -T{\cal S},
\eeq 
where the entropy density $\cal S$ is given by
\beq
{\cal S}= \sum_b {\cal S}_b+\sum_l {\cal S}_l
\eeq
with,
\bwt
\bea
{\cal S}^c_b&=&-\frac{|q_{b}|B}{2\pi^ {2}}\sum_{\nu, s}\int^{\infty}_{0}
dk^b_{\parallel}\left\lbrace f^b_{k,\nu, s}\log f^b_{k,\nu, s}+(1-f^b_{k,\nu, s})\log(1- 
f^b_{k,\nu, s})+\bar{f}^b_{k,\nu, s}\log\bar{f}^b_{k,\nu, s}+(1-\bar{f}^b_{k,\nu, s})\log(1-
\bar{f}^b_{k, \nu, s})\right\rbrace \cr
{\cal S}^n_b&=& -\frac{1}{2\pi^ {2}}\sum_{s}\int^{\infty}_{0}k^{b}_{\bot}
dk^{b}_{\bot} \int^{\infty}_{0} dk^{b}_{\parallel} 
\left\lbrace f^b_{k, s}\log f^b_{k, s}+(1-f^b_{k, s})\log(1- f^b_{k, s})+
\bar{f}^b_{k, s}\log\bar{f}^b_{k, s}+(1-\bar{f}^b_{k, s})\log(1-
\bar{f}^b_{k, s}) \right\rbrace \cr
{\cal S}_l &=&-\frac{|q_{l}|B}{2\pi^ {2}}\sum_{\nu, s}\int^{\infty}_{0}
dk^{l}_{\parallel}\left\lbrace f^l_{k, \nu, s}\log f^l_{k, \nu, s}+(1-f^l_{k, \nu, s})
\log(1- f^l_{k, \nu, s})+\bar{f}^\nu_{k, \nu, s}\log\bar{f}^\nu_{k, \nu, s}+(1-\bar{f}^\nu_{k, \nu, s})
\log(1-\bar{f}^\nu_{k, \nu, s}) \right\rbrace\nonumber\\
\eea
\ewt
The pressure of neutron star matter
is given by 
\beq
P_{m}=- \Omega=\mu_{n}\sum_{b}
\rho^{v}_{b} -\varepsilon_{m}+ T{\cal S},
\label{press}
\eeq
where the charge neutrality and $\beta$-equilibrium conditions are used to 
get the last equality. 
If the stellar matter contains neutrinos trapped, 
their energy density, pressure, and entropy contributions, respectively,
\bea
\varepsilon_{\nu_e}&=&\frac{1}{2\pi^ {2}} \int^{\infty}_{0}k^3 dk 
\left\lbrace f^\nu_{k, s}+\bar{f}^\nu_{k, s}\right\rbrace \cr
P_{\nu_e}&=&\frac{1}{6\pi^ {2}} \int^{\infty}_{0}k^3 dk \left\lbrace 
f^\nu_{k, s}+\bar{f}^\nu_{k, s}\right\rbrace\cr
{\cal S}^{\nu_e}_l &=&-\frac{1}{2\pi^ {2}} \int^{\infty}_{0}(k^{l})^2 dk^{l} 
\left\lbrace f^\nu_{k, s}\log f^\nu_{k, s}+(1-f^\nu_{k, s})\log(1- f^\nu_{k, s}) +
\bar{f}^\nu_{k, s}\log\bar{f}^\nu_{k, s}+(1-\bar{f}^\nu_{k, s})\log(1-\bar{f}^\nu_{k, s}) 
\right\rbrace,
\eea
should be added to the stellar matter energy and pressure.

\section{Results and discussion}

\begin{figure}[th]
\vspace{1.5cm}
\centering
\includegraphics[width=0.75\linewidth,angle=0]{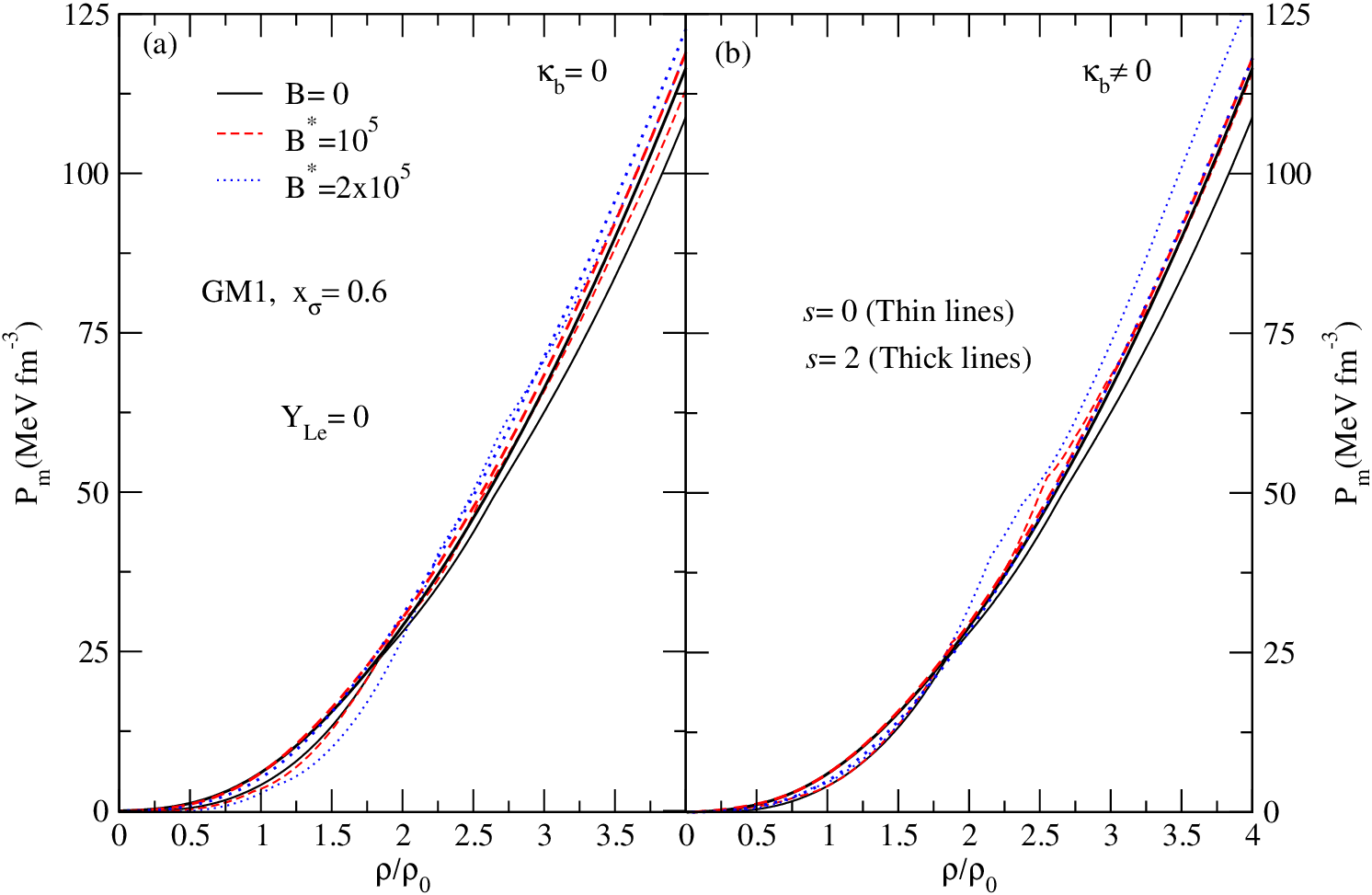}\\
\caption{(Color online) Matter pressure  as a function of the baryonic density, for several values of magnetic field ($B^*=0$, $B^{*}=10^5,\; 2\times 10^5$) without (left panels) and with AMM (right panels) for an entropy per baryon ${\mathfrak{s}}=0$, and 2, and for neutrino free matter.}
\label{eosnf}
\end{figure}

\begin{figure}[th]
\vspace{1.5cm}
\centering
\includegraphics[width=0.75\linewidth,angle=0]{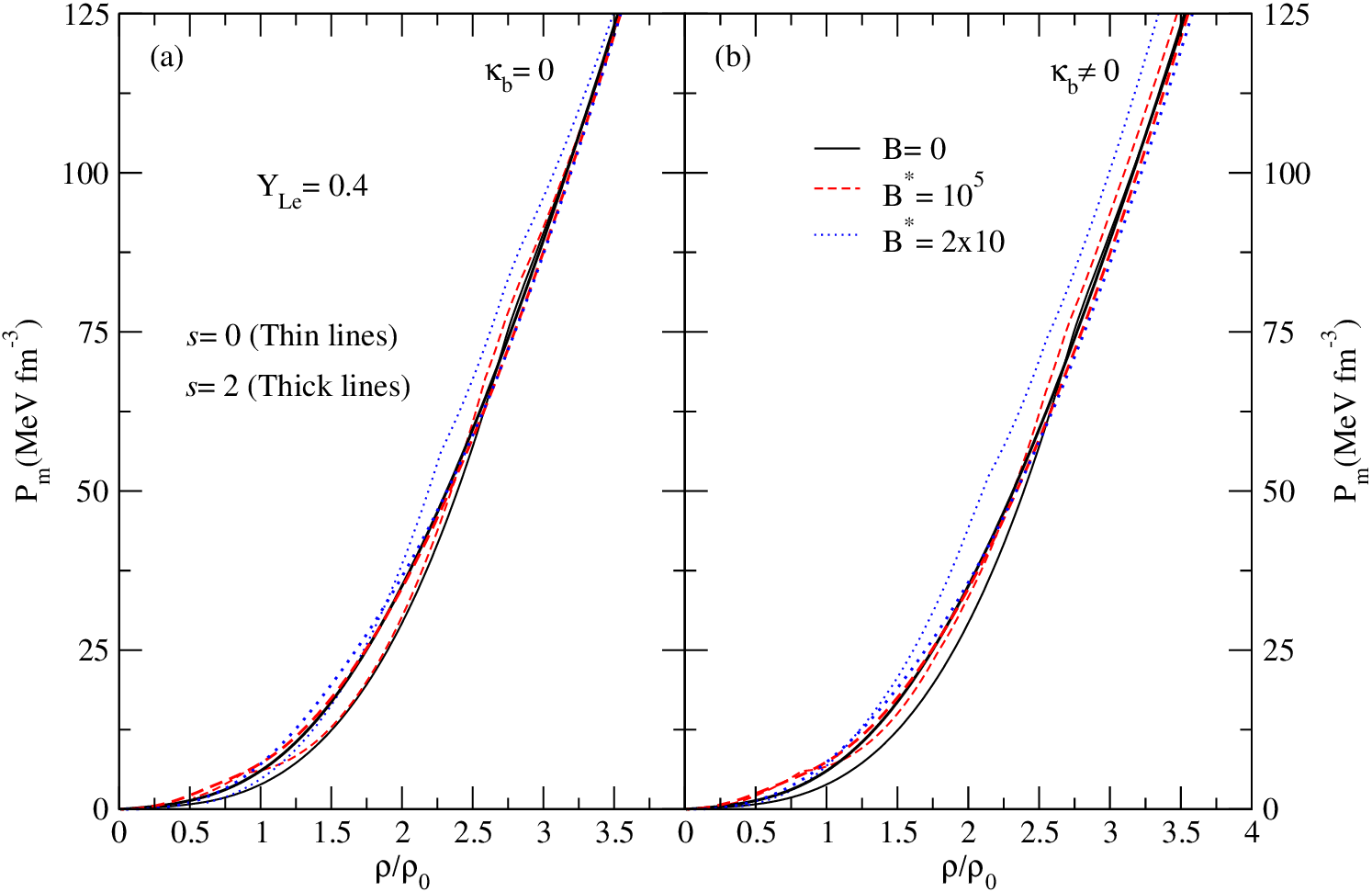}
\caption{(Color online) The same as Fig.~\ref{eosnf} but for matter with trapped neutrinos and lepton fraction $Y_{Le}=0.4$.}
\label{eosnt}
\end{figure}

We now study the stellar matter at finite temperature with magnetic fields.
We include the baryonic octet in the EOS and choose the  GM1 parameter set~\cite{gm91} for
our calculation. The static properties of the baryons were given in previous works~\cite{broderick,aziz09}. The parameters of the model are the nucleon mass 
$m_n=939$ MeV, the masses of mesons $m_\sigma$, $m_\omega$, $m_\rho$ and the 
coupling constants. 
The meson-hyperon couplings are assumed to be fixed fractions of the
meson-nucleon couplings, $g_{i H}=x_{i H} g_{i N}$, where for each meson $i$,
the values of $x_{i H}$ are assumed equal for all hyperons $H$. The values of
$x_{i H}$ are chosen to reproduce the binding energy of the $\Lambda$ at
nuclear saturation as suggested in~\cite{gm91}, 
and given in Table~\ref{table2}. A different choice could have been done in order to consider
that the optical potential  of the $\Sigma^-$ in nuclear matter is repulsive as was done in~\cite{chiap09}. However, there is very little experimental information that can be used to fix
the $\Sigma^-$ interaction. More, the main difference that would occur would be the onset of the $\Sigma^-$ at larger densities and the $\Xi^-$ at lower densities.
 
We consider that the external magnetic field is constant. The magnetic field 
will be defined in units of the critical field $B_e^c=4.414\times 10^{13}$ G so
that $B=B^*B_e^c$.
\begin{table*}
\caption{The parameter set GM1 \cite{gm91}used in the calculation.}
\label{table2}
\begin{ruledtabular}
\begin{tabular}{ c c c c c c c c c c cc}
$\rho_{0}$& -B/A & &$g_{\sigma N}/m_{\sigma}$ &$g_{\omega N}/m_{\omega}$&$g_{\rho N}/m_{\rho}$&  &  &  & &\\
($fm^{-3}$) &(MeV)& $M^{*}/M$ & (fm) & (fm) & (fm) &$x_{\sigma H}$&$x_{\omega H}$ & $x_{\rho H}$ &b &c \\
\hline
0.153&16.30&0.70& 3.434 & 2.674 & 2.100 & 0.600 & 0.653 & 0.600 &0.002947 
&-0.001070 \\
\end{tabular}
\end{ruledtabular}
\end{table*}

In Fig.~\ref{eosnf} and Fig.~\ref{eosnt} the pressure is plotted versus density for $B^*=0,\, 10^5$ and $2\times 10^5$ for neutrino free matter (Fig.~\ref{eosnf}) and matter with trapped neutrinos (Fig.~\ref{eosnt}), without and with the inclusion of the AMM, respectively, in left and right panels. We consider an entropy per baryon ${\mathfrak{s}}=0$ and 2. The kink on each EOS curve identifies the onset of hyperons. The effects of the AMM are only noticeable
for a  strong magnetic field  above $B^*=10^5$ as already discussed in~\cite{broderick}.

\begin{figure}
\vspace{1.5cm}
\centering
\includegraphics[width=0.75\linewidth,angle=0]{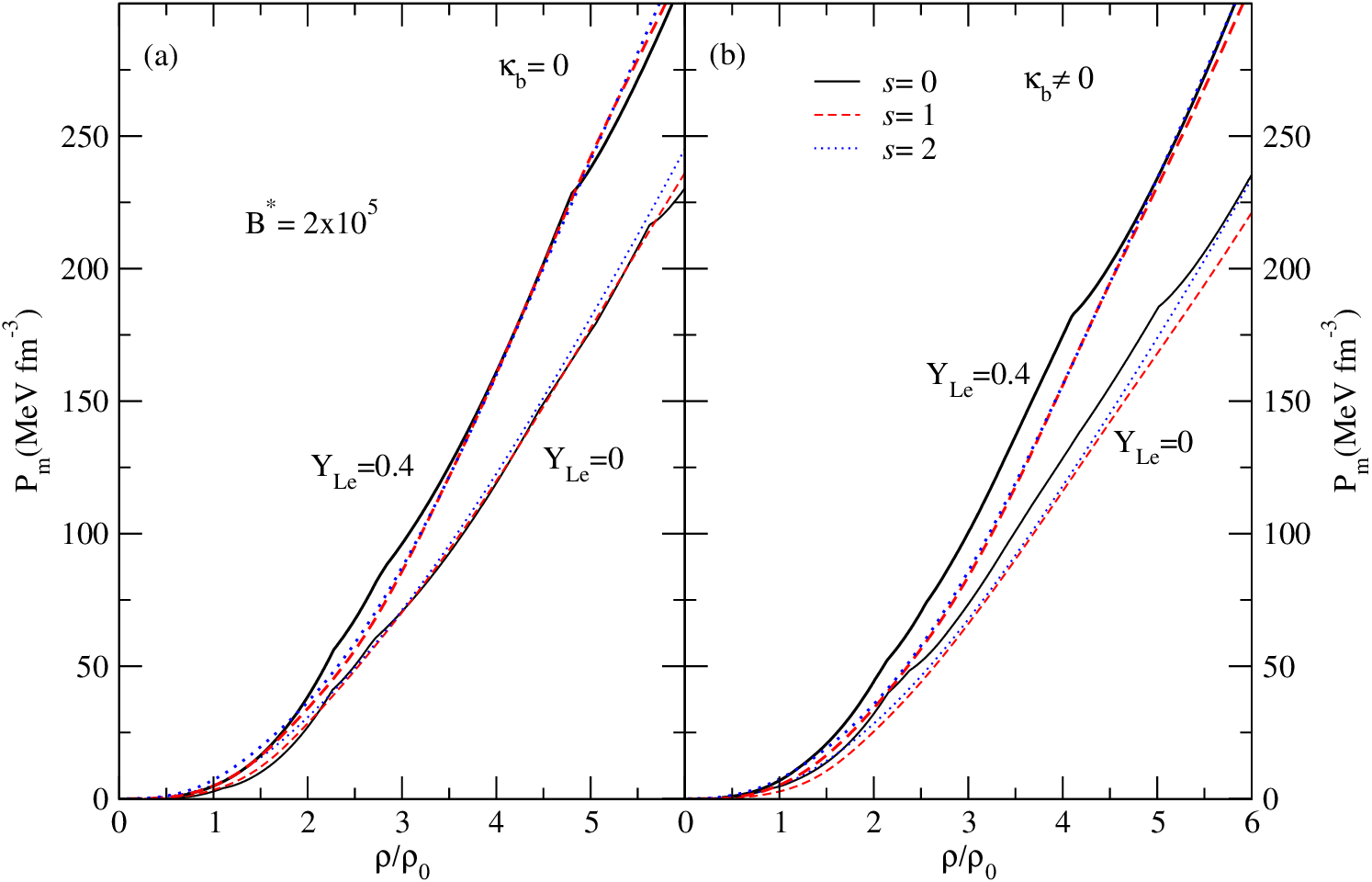}\\
\caption{(Color online) Matter pressure  as a function of the baryonic density, for $B^*=2\times 10^5$: comparison between both neutrino free and neutrino trapped matter for several values of the entropy per baryon $\mathfrak{s}$.}
\label{eos3}
\end{figure}

We will start by discussing neutrino free matter, shown in Fig.~\ref{eosnf}.
 The strong magnetic field makes the EOS softer at lower
densities and harder at  higher densities due to the Landau quantization
which affects charged particles~\cite{broderick,aziz09}. However, at finite temperature these
effects are partially washed out and, we may see in the left panel of Fig. \ref{eosnf}  that the
EOS for $B^*=2\times 10^5$ is not much softer than the corresponding EOS for $B^*=0$ in the lower range of densities plotted. The effects of temperature are even stronger when the AMM is
included (right panel). As discussed in~\cite{broderick},  in the presence of a strong magnetic
field  the extra hardness due to the inclusion of AMM is mainly due to an increase of the
neutron pressure degeneracy due to spin polarization of the neutrons. Temperature raises
partially this degeneracy and, consequently, gives rise to a softening of the EOS: at large
densities  the three EOS plotted for ${\mathfrak{s}}=2$ with $B^*=0, 10^5, 2\times 10^5$ almost coincide. While for zero magnetic field the EOS becomes harder with the inclusion of
temperature, the opposite may occur for a strong magnetic field.  One of the main effects of
temperature on the EOS of matter under a strong magnetic field is to wash out the effects of the Landau quantization and spin polarization.

In Fig.~\ref{eosnt} we show results for matter with trapped
neutrinos and a lepton fraction $Y_{Le}=0.4$. In this case, at low densities, above $\rho=0.5\rho_0$, the EOS becomes harder for a finite external magnetic field. This is mainly  due to the larger electron fraction below the onset of strangeness.  However, just as before, the effect of the field is not so strong for a finite entropy per baryon. Above $\rho=2\rho_0$ the equation becomes softer in the presence of a magnetic field both taking into account AMM or not if ${\mathfrak{s}}=2$, although for ${\mathfrak{s}}=0$ the opposite occurs. This is due to the fact that above this density for $B^*=2\times 10^5$  the proton fraction becomes comparable or even larger than the neutron fraction, and therefore the kinetic contribution to the pressure reduces. When the AMM is included there is not much difference at ${\mathfrak{s}}=2$, however for ${\mathfrak{s}}=0$ the EOS is a slightly harder than in the case without AMM due to a larger electron fraction and smaller hyperon fraction.

In Fig.~\ref{eos3} we have plotted for $B^*=2\times 10^5 $ and ${\mathfrak{s}}=0,\, 1$ and 2, the pressure for neutrino free matter and neutrino trapped matter, without  (left panel) and with (right panel) AMM in order to compare the effect of a strong magnetic field in matter with and without neutrinos. At zero magnetic field the EOS of neutrino free matter becomes harder at finite temperature for the lower densities~\cite{prakash97}. However, the onset of strangeness occurs at lower densities for finite temperature and this will soften the EOS~\cite{prakash97,menezes03,panda10}. As a
consequence, for a range of intermediate densities it may occur that the EOS at finite
temperature is softer~\cite{prakash97,panda10}.  Trapped neutrinos increase (decrease) the proton (neutron) fraction at low density and this softens the EOS relatively to the neutrino free case due to an overall smaller baryonic  kinetic pressure. However, at larger densities trapped neutrinos hinder the onset of
hyperons, therefore shifting the softening of the EOS due to strangeness to larger densities.
 For $B^*=2\times 10^5$ and zero temperature neutrino trapped matter is not softer than neutrino free matter, because the decrease on the neutron fraction due to trapped  neutrinos is much smaller and the overall decrease on the baryonic kinetic pressure does not compensate the increase of the total lepton contribution to the EOS. At finite temperature, however, the onset of hyperons at lower densities and the crossing of the neutron with the proton fraction also at a lower density gives rise to a softening of the EOS for densities $2\rho_0<\rho<3\rho_0$.
The inclusion of the AMM makes the nucleonic fraction of neutrino trapped matter closer to the ones of neutrino free matter, and, therefore, the softening that occurs due to a more equilibrated distribution of protons and neutrons does not compensate the larger lepton contribution in neutrino trapped matter. However, finite temperature washes LL filling  effects and spin polarization effects and a softer EOS results.

\begin{figure}[ht]
\vspace{1.5cm}
\centering
\includegraphics[width=0.85\linewidth,angle=0]{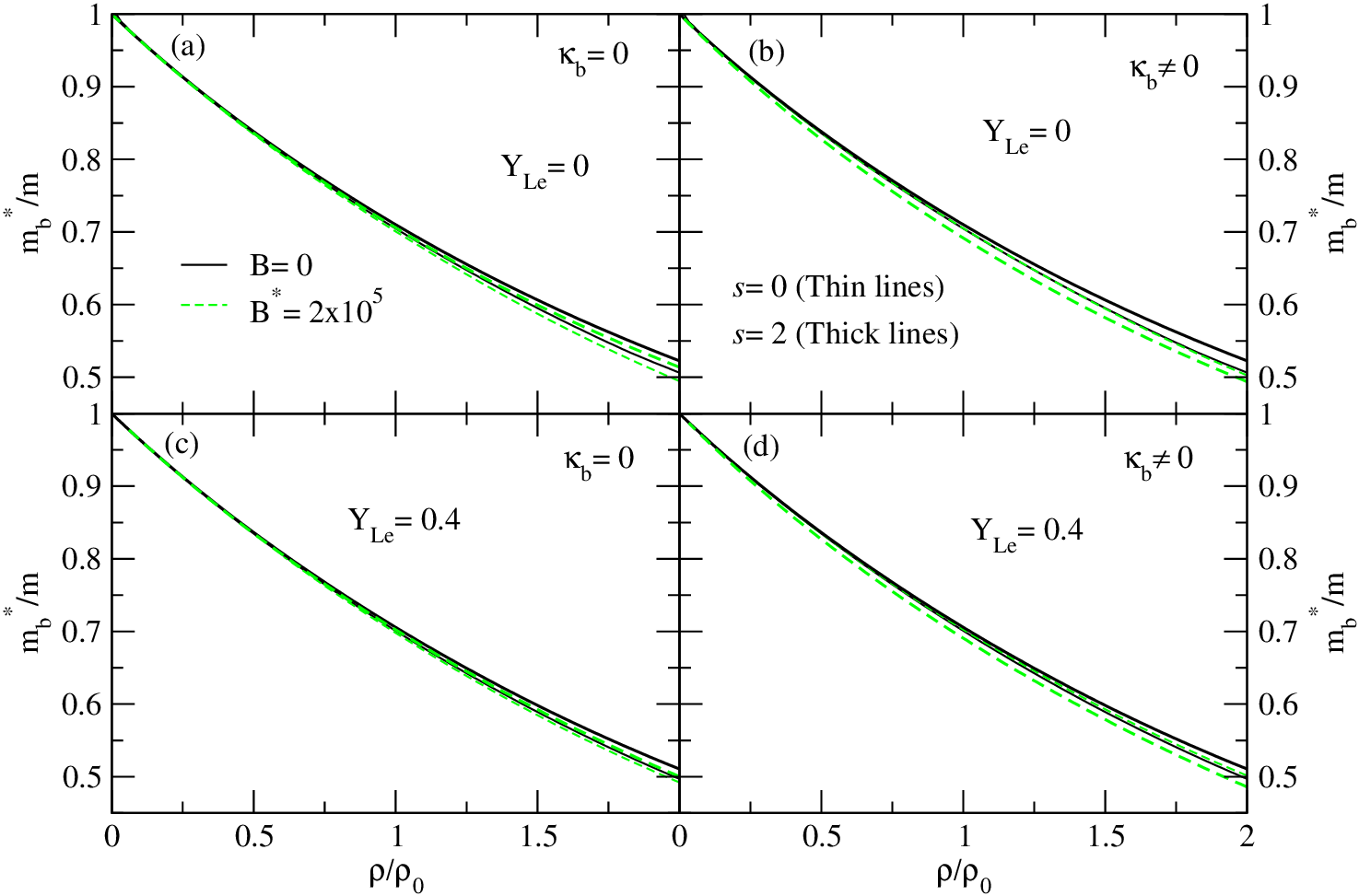}
\caption{(Color online) Nucleon effective mass  as a function of the baryonic density, for  $B^*=0,\; \hbox{and}\; 2\times 10^5$, without (left panels) and with AMM (right panels), for neutrino free matter (top panel) and matter with trapped neutrinos (bottom panel).}
\label{effe}
\end{figure}

In Fig.~\ref{effe}, the nucleon  effective mass is shown as function
of the baryon density for ${\mathfrak{s}}=0$ and  ${\mathfrak{s}}=2$, and for $B^*=0,\;
\hbox{and}\; 2\times 10^5$ without and with AMM. Figs.~\ref{effe} (a), (b)  correspond to
neutrino free matter, and Figs.~\ref{effe} (c) and (d) to matter with trapped neutrinos and a
lepton fraction $Y_{Le}=0.4$. For ${\mathfrak{s}}=0$,  if the AMM is not taken into account,  the
effective mass reduces faster with an increase of the density at finite magnetic field
 while the opposite occurs when AMM is included,~\cite{broderick}.
However, at finite temperature, even taking into account the AMM, the nucleon effective mass
decreases at a given density when $B^*$ increases because spin polarization effects are
partially washed out with temperature. We will see later that this has an important
effect on the temperature of matter with a fixed entropy per baryon.

\begin{figure}[ht]
\vspace{1.5cm}
\centering
\begin{tabular}{c}
\includegraphics[width=0.85\linewidth,angle=0]{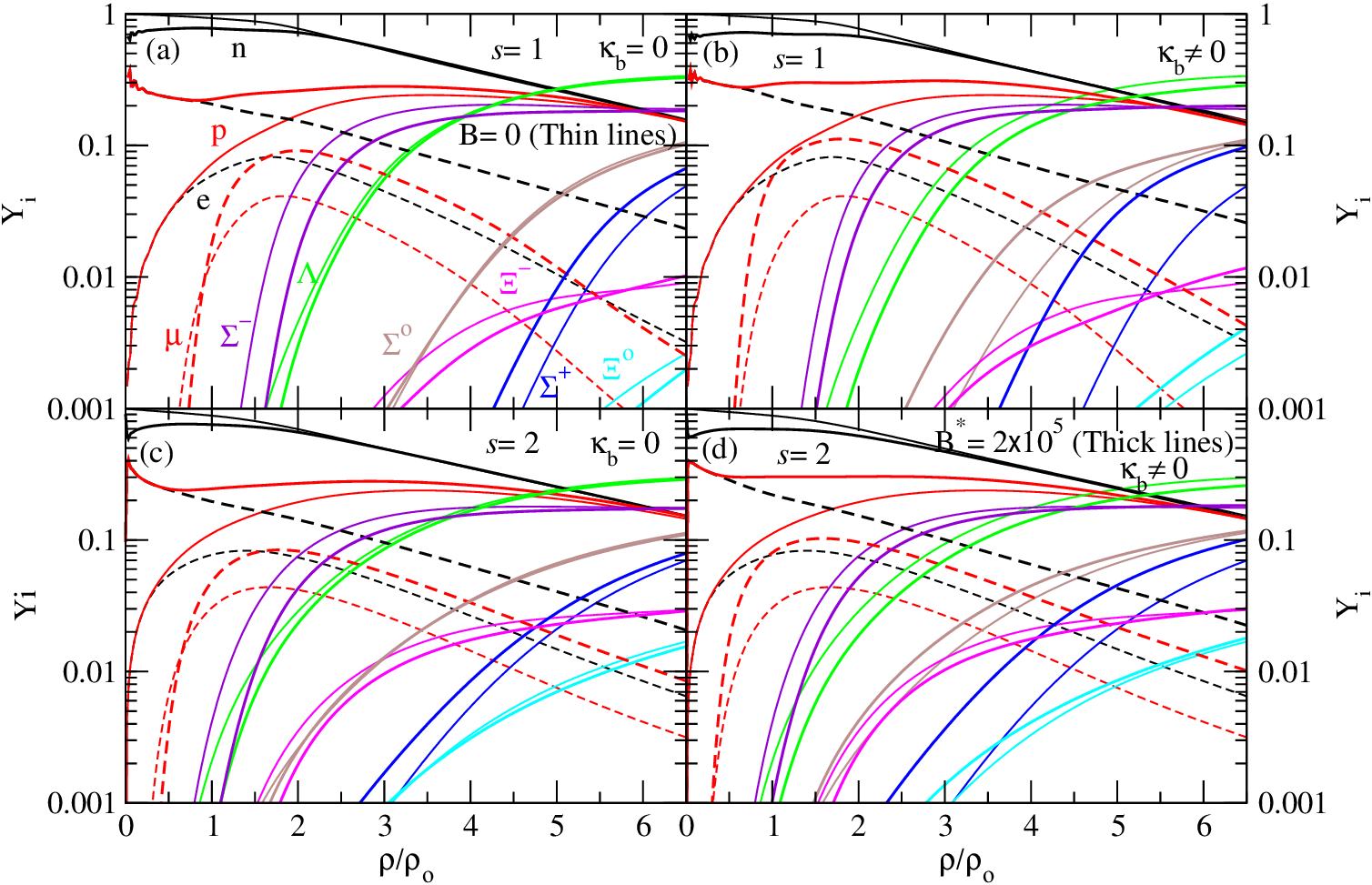}\\
\vspace{1.5cm}\\
\includegraphics[width=0.85\linewidth,angle=0]{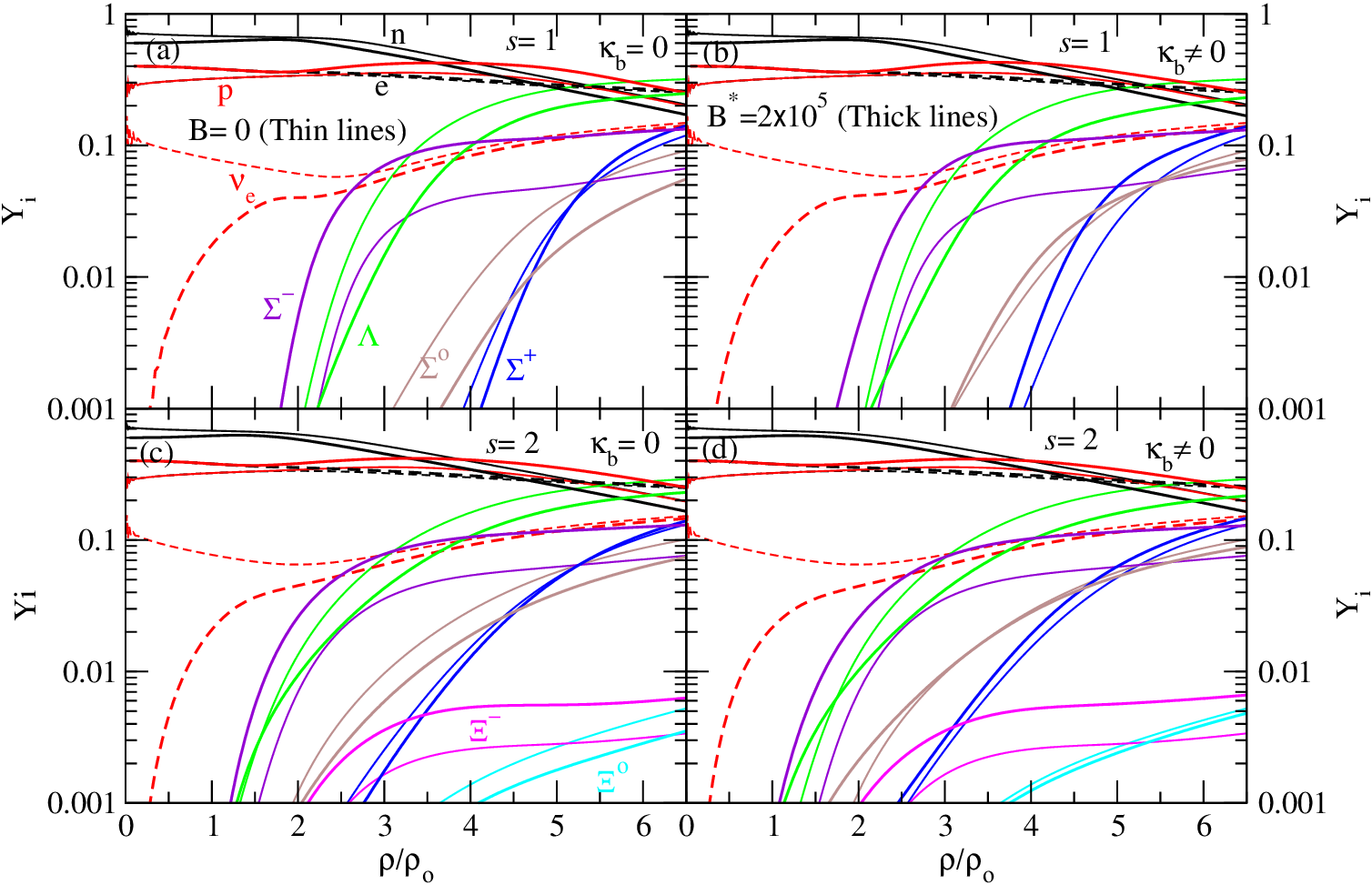}
\end{tabular}
\caption{(Color online) Particle fractions   as a function of the baryonic density, for  $B^*=0$ and $2\times 10^5$, without (left panels) and with  AMM (right panels), for neutrino free matter (top panel) and matter with trapped neutrinos
(bottom panel). }
\label{frac1}
\end{figure}

In Fig.~\ref{frac1} the  particle fraction $Y_i=\rho_i/\rho$ for baryons and leptons  is plotted as a function of
the baryon density for $B^*=2\times 10^5$ (thick lines) and $B^*=0$ (thin lines) for two values of
the entropy per particle, ${\mathfrak{s}}=1$ and 2.
 Neutrino free matter is represented in the top panel
and matter with trapped neutrinos in the bottom panel. At zero magnetic field, the main effect
of temperature is to move the onset of hyperons to lower densites~\cite{prakash97}. This
feature is still true for a finite magnetic field.
 If matter is under the effect of a strong
magnetic field we find the same effects at finite temperature as at zero temperature: a) if the
AMM is not included the onset
of the $\Sigma^-$  is shifted to larger densities, the onset of $\Sigma^+$ to smaller densities,
and the neutral hyperons are not much affected; b) including the AMM  the onset $\Sigma^-$  is
shifted to larger densities but less than in the previous case, the one of $\Sigma^+$ is
shifted to even  smaller densities,
and in this case the neutral hyperons are  affected with $\Lambda$ behaving like $\Sigma^-$ and
 $\Sigma^0$ like  $\Sigma^+$. The onset of the  cascate $\Xi^-$ occurs below $6\rho_0$ for
${\mathfrak{s}}=1$ and 2, and behaves like $\Sigma^-$ with the magnetic field.

 As discussed in~\cite{broderick1,aziz10} the behaviour observed among the different hyperons
is mainly due to a decrease of the neutron chemical potential, due to a smaller isospin
asymmetry, and a decrease of the electron chemical potential due to Landau quantization.
In fact, at low
densities, the fraction of proton and leptons are significantly affected by
the magnetic field. The Landau quantization increases the proton abundance,
and, therefore, the electron abundance due to the charge neutrality condition. 
 The inclusion of AMM reduces the chemical potential of all the hyperons and a
complicated balance between the different terms, including the magnitude of the AMM, will
define whether the onset is shifted to larger or smaller densities. For instance the AMM of
$\Sigma^0$,  $\Sigma^+$  is much larger than the one of $\Sigma^-$, $\Lambda$ and $\Xi^-$
(see Fig.~\ref{frac1} bottom panel) and this may
explain the different behavior of these hyperons when AMM is included.

At $B=0$ and $T=0$, the presence of neutrinos shifts the
onset of hyperons to larger densities because the neutron chemical potential is smaller and the
neutrino chemical potential finite, with temperature making this effect less effective. For
warm matter under a strong magnetic field we conclude: a) if AMM is not taken into account, a
smaller chemical potential of both neutrons and neutrinos explains a smaller (larger) shift of the negatively
(positively) charged baryons to larger densities, and  a
smaller neutron chemical potential gives rise to a shift to larger densites of the neutral
baryons onset; b) including AMM may change these  trends mainly the ones associated with the baryons with a larger AMM, because the hyperon chemical potentials decrease and, therefore the onset of all hyperons will occur at smaller densities than in the absence of AMM.

\begin{figure}[ht]
\vspace{1.5cm}
\centering
\begin{tabular}{c}
\includegraphics[width=0.8\linewidth,angle=0]{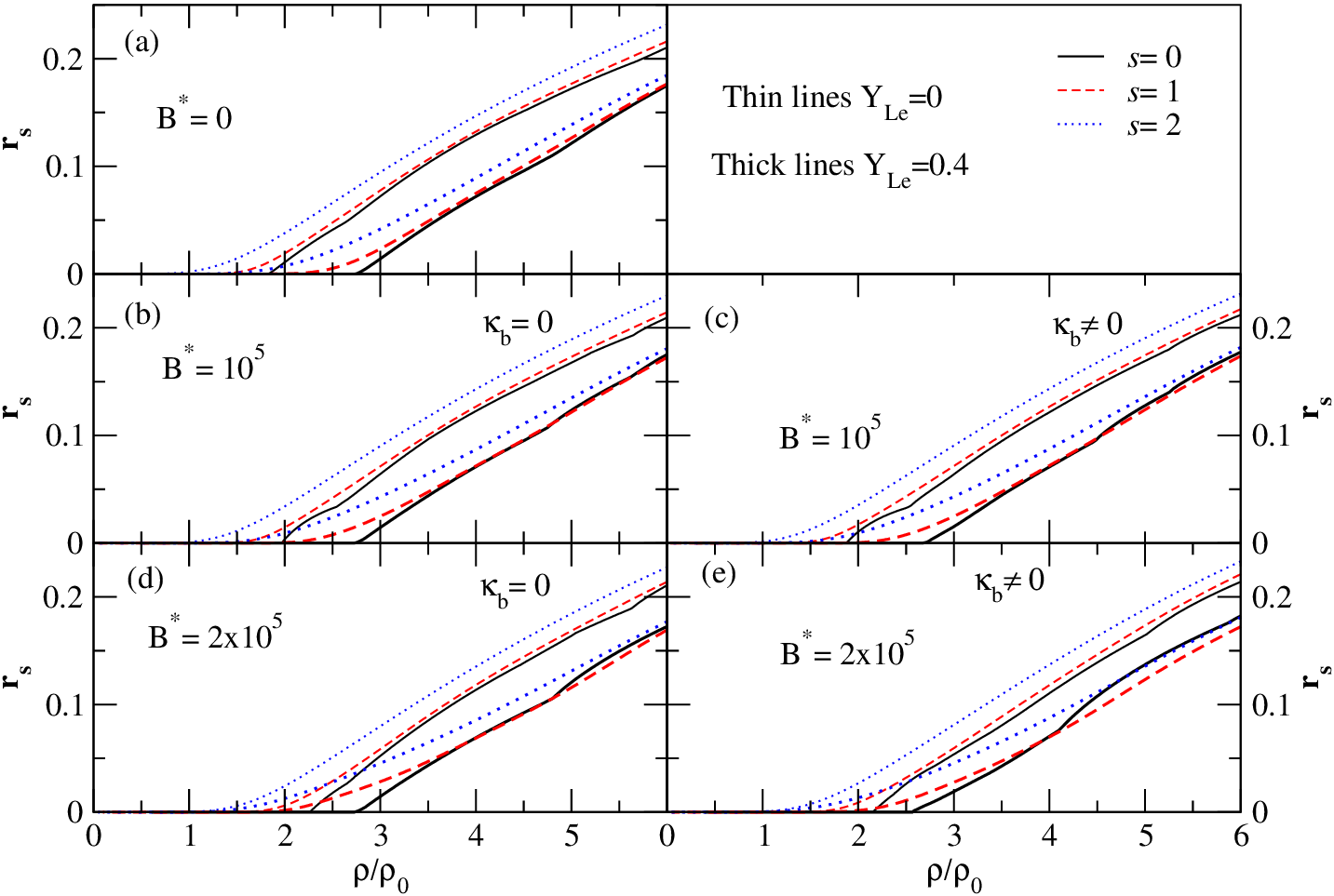}\\
\vspace{1.5cm}\\
\includegraphics[width=0.8\linewidth,angle=0]{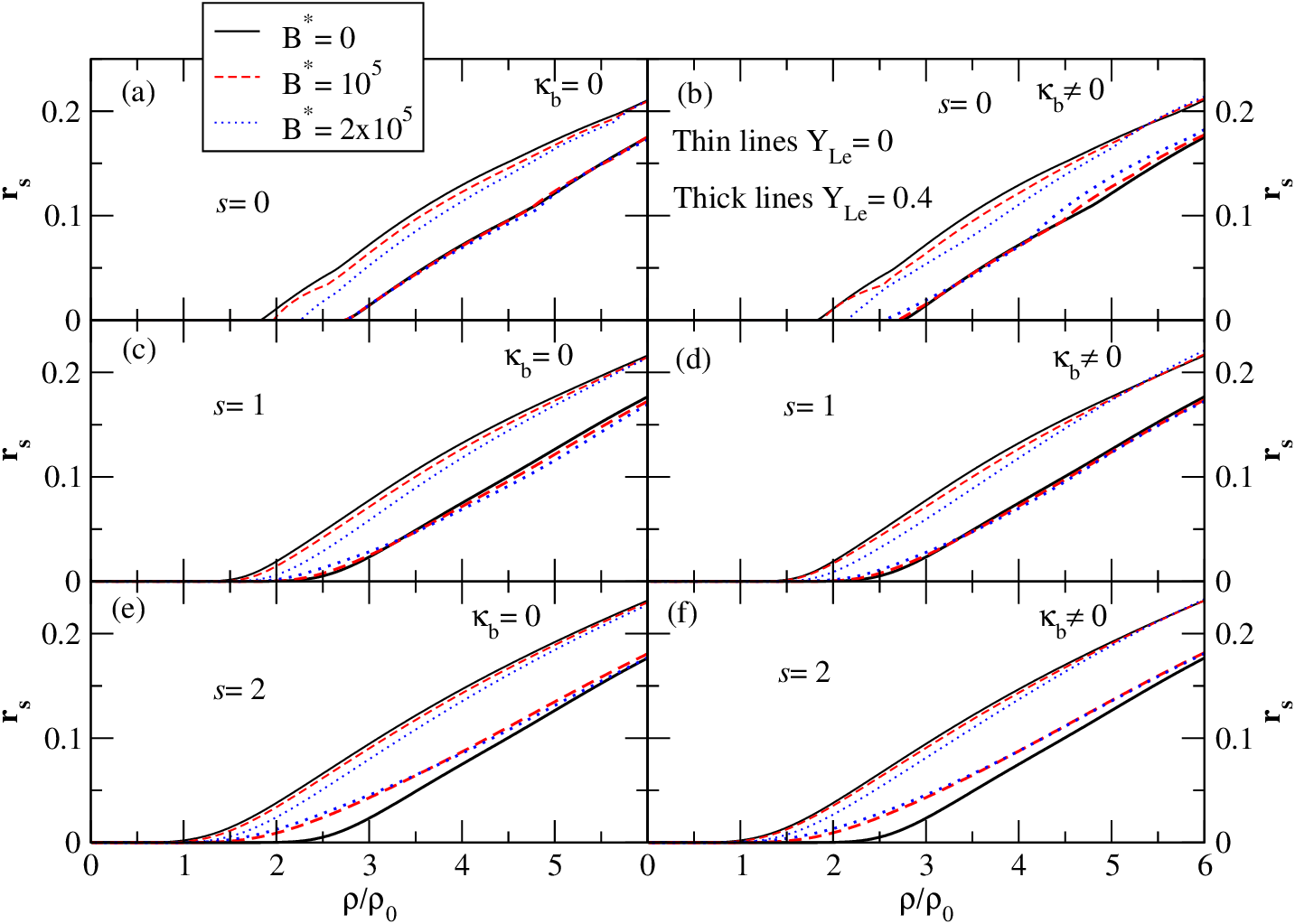}
\end{tabular}
\caption{(Color online) Strangeness fraction  as a function of the baryonic density, for 
severals values of the magnetic field B and ${\mathfrak{s}}=0,1,2$, without and with AMM, for GM1 model. Top panel: the strangeness fraction at a fixed entropy for different values of $B$; bottom panel: the strangeness fraction at a fixed $B$ for different values of the entropy per particle ${\mathfrak{s}}$.}
\label{strange1}
\end{figure}

The effect of the magnetic field on the total strangeness fraction is better seen in Fig.~\ref{strange1}, where we show the total strangeness fraction for different values of magnetic field and different values of the entropy per baryon $\mathfrak{s}$ with and without AMM. The fraction of the strangeness in the system is given by
\begin{equation}
r_{\mathfrak{s}}=\frac{\sum_bq_s^b\rho_b}{3\rho},
\end{equation}
where $q_s^b$ is the strange charge of baryon $b$.
At zero entropy the  strangeness onset occurs respectively, below $2\rho_0$ and   $3\rho_0$
for neutrino free and neutrino  trapped matter. At $\rho=6\rho_0$ neutrino trapped matter has a strangeness fraction 0.03-0.04 smaller due to the larger electron fraction. The magnetic field lowers the strangeness fraction for neutrino free matter and almost does not affect the strangeness fraction for neutrino trapped matter.  This effect  was already discussed in \cite{aziz10} and is due to the fact that with the presence of neutrinos the neutron chemical potential does not suffer such a large reduction as in neutrino free matter.
The top panel of Fig.~\ref{strange1} allows a comparison of the strangeness fraction at a fixed entropy and different $B$ intensities.
 For a finite entropy the effect of $B$ in neutrino free matter is equivalent to the one already obtained at ${\mathfrak{s}}=0$, there is a reduction of strangeness with the increase of $B$. However, in neutrino trapped matter the opposite may occur and  for ${\mathfrak{s}}=2$  the larger the magnetic field  the larger is strangeness fraction. This is true whether AMM is taken into account or not.  
In neutrino free matter the effect of temperature is not strong enough to oppose the shift of the onset of strangeness to larger densities due to the decrease of the neutron chemical potential.

In the bottom panel of Fig.~\ref{strange1} we compare the strangeness fraction for a fixed magnetic field and different entropies. Except for the larger field it is verified the general trend of temperature that shifts the onset of strangeness to lower densities for both neutrino trapped and neutrino free matter~\cite{menezes04,panda10}. However, in neutrino trapped matter at large densities this trend may fail if the temperature is not high enough.

\begin{figure}[ht]
\vspace{1.5cm}
\centering
\includegraphics[width=0.6\linewidth,angle=0]{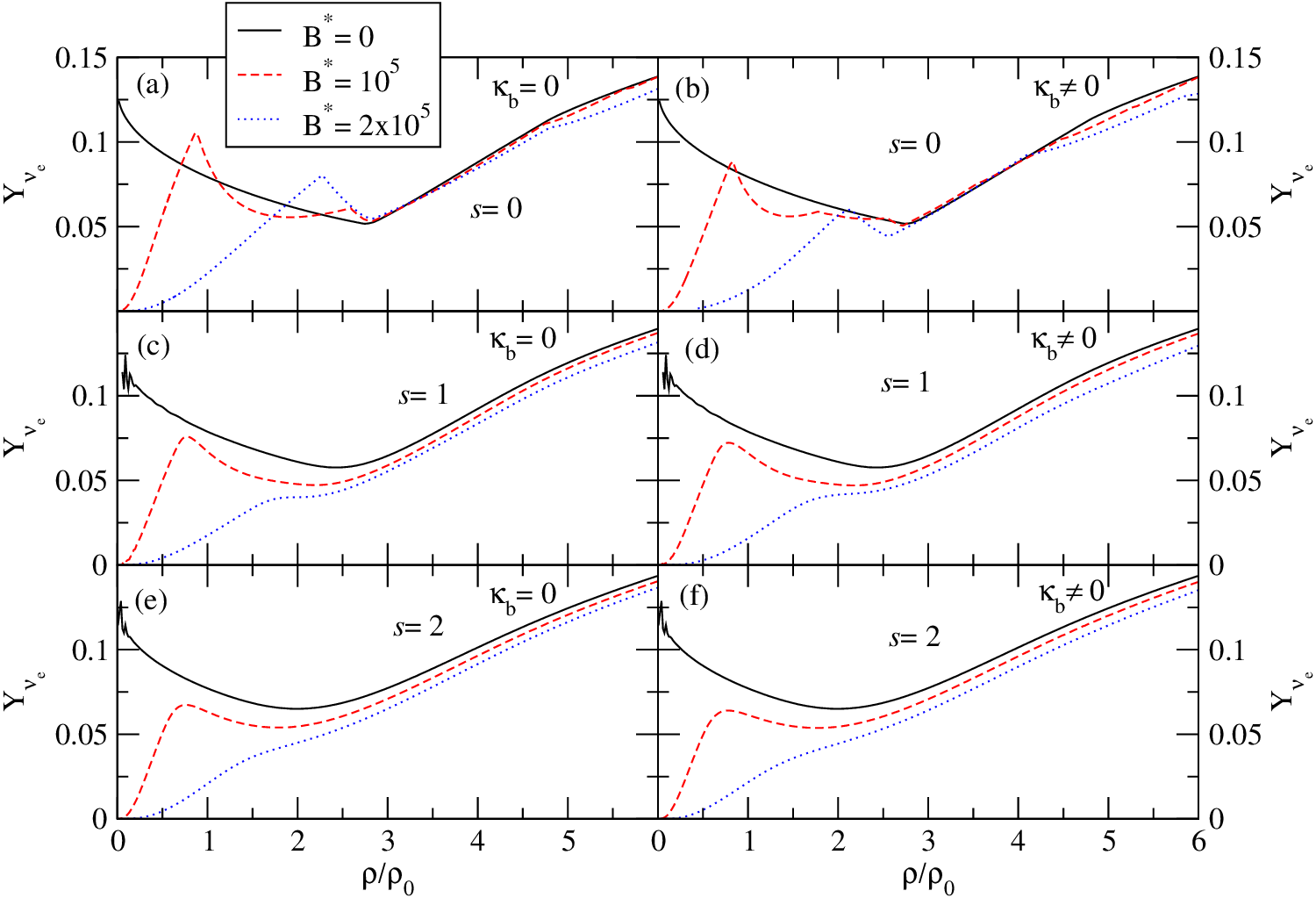}
\caption{(Color online) Neutrino fraction  as a function of the baryonic density, for 
severals values of of the magnetic field B and entropy per baryon ${\mathfrak{s}}$, without (left panel)  and with (right panel)  AMM.}
\label{fneutrino}
\end{figure}

It is also interesting to analyse the effect of $B$ on the neutrino fraction.
shown in Fig. \ref{fneutrino} as a function
of the baryonic density for several values of magnetic field and the entropy.
 In \cite{aziz10} it was discussed that, for zero temperature, the magnetic field
gives rise to a strong neutrino suppression at small densities, as seen in the top panels. This
was attributed to the large proton and, therefore, also  electron fractions. At finite
temperature the suppression at low densities persists, although, the fluctuations due to the
filling of the Landau levels disappear. It is  seen that for a finite entropy also at high
densities a strong magnetic field gives rise to a decrease of neutrino fraction,  due to the larger proton fraction that favors a larger electron fraction. A smaller neutrino fraction may imply a slower cooling of the star core, since the core cools essentially by neutrino emission \cite{prakash97}.

\begin{figure}[ht]
\vspace{1.5cm}
\centering
\includegraphics[width=0.75\linewidth,angle=0]{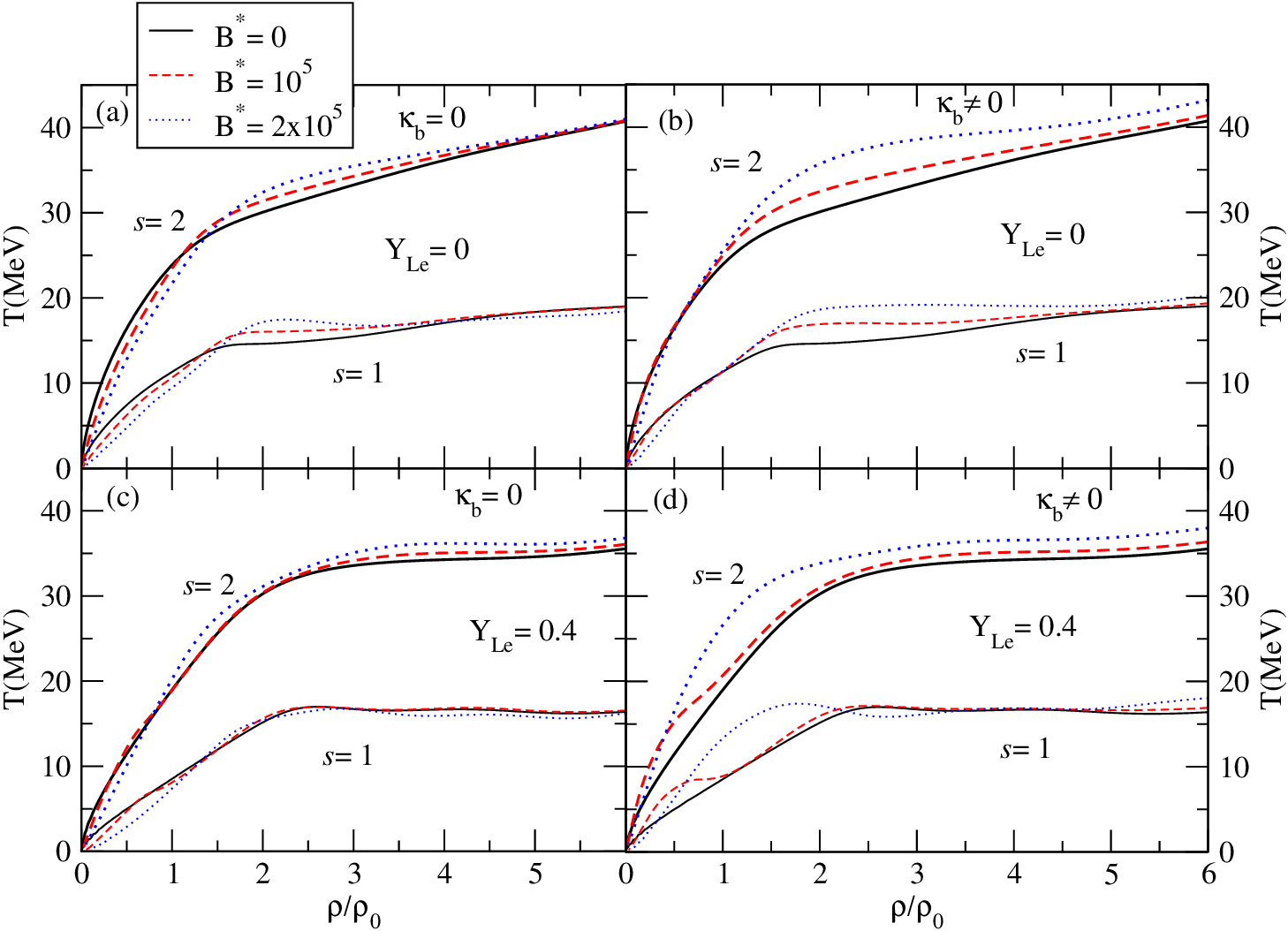}
\caption{(Color online) Temperature as a function of the baryonic density, for severals 
values of of the magnetic field B, without (left panels) and with (right panels) AMM for neutrino free matter (top panels) and matter with trapped neutrinos (bottom panels).
The thin line is for ${\mathfrak{s}}=1$ and the thick line for ${\mathfrak{s}}=2$.}
\label{temp1}
\end{figure}

In Fig.~\ref{temp1}, the temperature of the system is plotted for a  entropy
per baryon ${\mathfrak{s}}=1$ and $2$, respectively, for neutrino free matter and matter with trapped neutrinos. As
expected, for a larger  entropy per baryon, higher temperatures are reached. 
In a strong magnetic field these temperatures can be even larger.
At low densities, before the onset of  hyperons, temperature rises slower for the larger
magnetic fields because the proton (neutron) fraction increases (decreases) with $B$ and the
nucleonic degrees of freedom become more equally distributed. The kink in all the curves
identifies the onset of hyperons, and as discussed before, occurs at lower densities for lower
values of $B$.  This is true both for ${\mathfrak{s}}=1$ and 2, with or without AMM. At high densities the
temperature becomes approximately constant: this is clearly seen for ${\mathfrak{s}}=1$; for ${\mathfrak{s}}=2$ the
temperature saturation occurs at higher densities because the hyperon fractions increase during
a larger range of densities until they attain a saturation fraction. If AMM is included the
temperature raises to larger values at finite $B$ probabily due to the larger lepton fraction. 

For matter with trapped neutrinos,  we have a similar situation. However,
because the proton fraction is larger due to the  fixed lepton fraction constraint, the effect
of $B$  in reducing the neutron fraction at low densities is not so large and  the dependence of
$T$ on the baryonic density does not depend on $B$ until the onset of hyperons, which occurs in a smaller
range of densities in matter with trapped neutrinos. Also a smaller overall neutron fraction
favors lower temperatures as compared with neutrino free matter. The differences occuring below
$2\rho$ for the description with AMM is mainly due to the increase of the neutrino fraction and a faster decrease of the effective mass with the baryonic density at finite temperature and finite $B$.

Since, to date, there is no information available on the interior magnetic 
field of the star, we will assume that the magnetic field is baryon 
density-dependent as suggested by Ref.~\cite{chakrabarty}. The variation 
of the magnetic field $B$ with the
baryons density $\rho$ from the center to the surface of a star is
parametrized~\cite{chakrabarty, Mao03} by the following form
\beq
B\left(\frac{\rho}{\rho_0}\right) =B^{\hbox{surf}} + B_0\left[1-
\exp\left\lbrace-\beta\left( \frac{\rho}{\rho_0}\right)^\gamma  
\right\rbrace\right],
\label{brho}
\eeq 
where $\rho_0$ is the saturation density, $B^{\hbox{surf}}$ is the magnetic
field at the surface taken equal to $10^{15}$G in accordance with the
values inferred from observations and $B_0$ represents the magnetic field
at large densities. The parameters, $\beta $ and $\gamma$ are
chosen in such way that the field decreases with the density from
the centre to the surface. In this work, we will use the set of values 
$\beta=0.05$ and $\gamma=2$, allowing a slowly varying field with the density.
The magnetic field will be considered in units of the critical field 
$B^c_e=4.414 \times 10^{13}$~G, so that $B_0=B^*_0 \, B^c_e$. We 
take $B^*_0$ as a free parameter 
to check the effect of magnetic fields on stellar matter. Here, we consider
the EOS by taking the hyperon-meson coupling constants $x_\sigma=0.6$.

Hadron star properties are obtained from the EOS studied, for several values of
magnetic field, by solving the Tolman-Oppenheimer-Volkoff equations resulting
from the Einstein's general relativity equations for spherically symmetric
static stars. This is as approximation since the magnetic field distroys the spherical
symmetry, and, therefore,  we  interpret the obtained results as average values. We do not
allow the magnetic field at the centre of the star to exceed $\sim 3\times 10^{18}$ G according
to the results of~\cite{broderick1} which indicate that stable stars do not occur with a larger
central magnetic field.

\begin{figure}[ht]
\vspace{1.5cm}
\centering
\begin{tabular}{ccc}
\includegraphics[width=0.45\linewidth,angle=0]{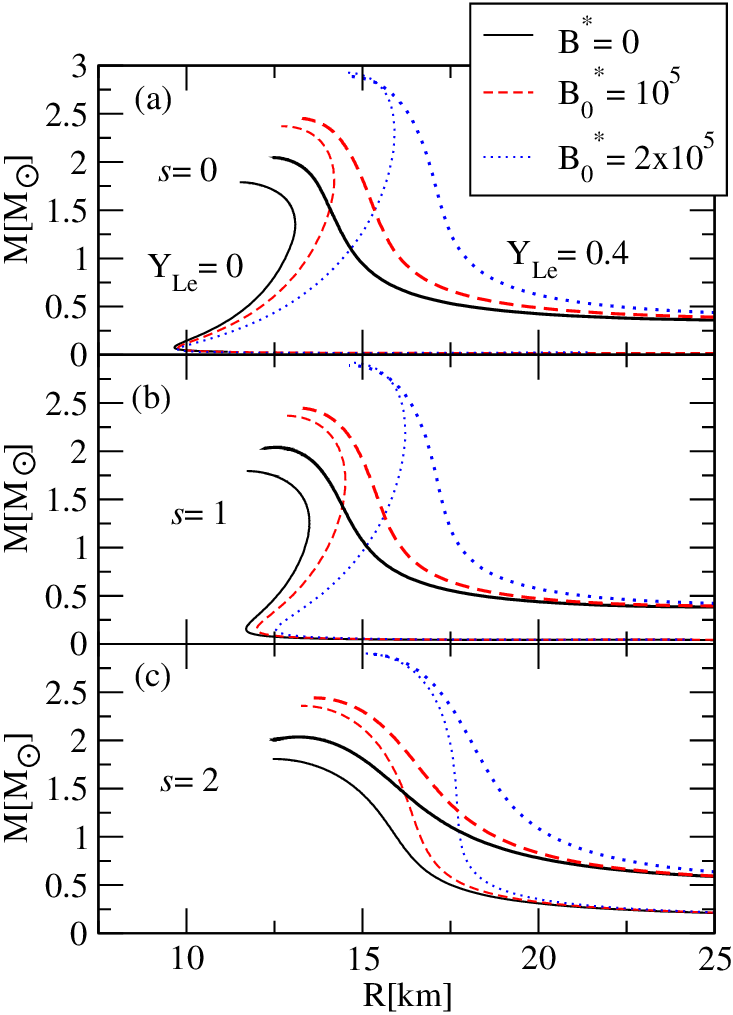}
\end{tabular}
\caption{(Color online) Mass-radius curve of neutron stars for several values of the magnetic
field, using a density dependent magnetic field $B$ given by Eq.~(\ref{brho}). Thin lines correspond to neutrino-free matter and thick lines to trapped neutrino matter with lepton fraction $Y_{Le}= 0.4$.}
\label{massgrav1}
\end{figure}

In Fig.~\ref{massgrav1} we show the family of stars
corresponding to the maximum mass configuration given in Tables~\ref{table3} and~\ref{table4}. Two main conclusions may be drawn a) warm stars with trapped neutrinos have larger
masses and radii. However, the differences get smaller for the most massive stars if a strong
magnetic field exists; b) a strong
magnetic field makes the star radius larger, as well as the  mass of the maximum mass star configuration.

 In table~\ref{table3}  the maximum 
gravitational and baryonic masses of stable stars, their radii, central
energy densities and magnetic field at the center, are given for neutrino free matter. In table \ref{table4} the same quantities are displayed for matter with trapped neutrinos.
 
The main conclusions we draw from the tables are:
a) the maximum baryonic mass of the star always decreases with ${\mathfrak{s}}$, independently of the existence of a magnetic field or trapped neutrinos;
b) for a finite magnetic field the maximum gravitational mass decreases slightly with ${\mathfrak{s}}$, however, the opposite occurs for $B=0$;
c) in the presence of a huge magnetic field the central baryonic density of the star is smaller and the radius larger.

Neutrinos diffuse out of the core after a first period when they are trapped in the core of a proto-neutron, and the star reachs an entropy per particle of ${\mathfrak{s}}\sim 1$ (in units of the Boltzmann constant). During the deleptonization period the core is heated up and reaches an entropy per particle  ${\mathfrak{s}}\sim 2$. After the core deleptonizes exotic degrees of freedoms such as hyperons will appear.
We will discuss how the magnetic field may influence the evolution of the star from a stage with ${\mathfrak{s}}=1$ and trapped neutrinos, to a stage  of warm neutrino free matter with ${\mathfrak{s}}=2$ and finally a cold neutrino free star with ${\mathfrak{s}}=0$. During this evolution the gravitational mass of the star decreases but its baryonic mass will stay constant.
 
\begin{figure}[ht]
\vspace{1.5cm}
\centering
\begin{tabular}{ccc}
\includegraphics[width=0.75\linewidth,angle=0]{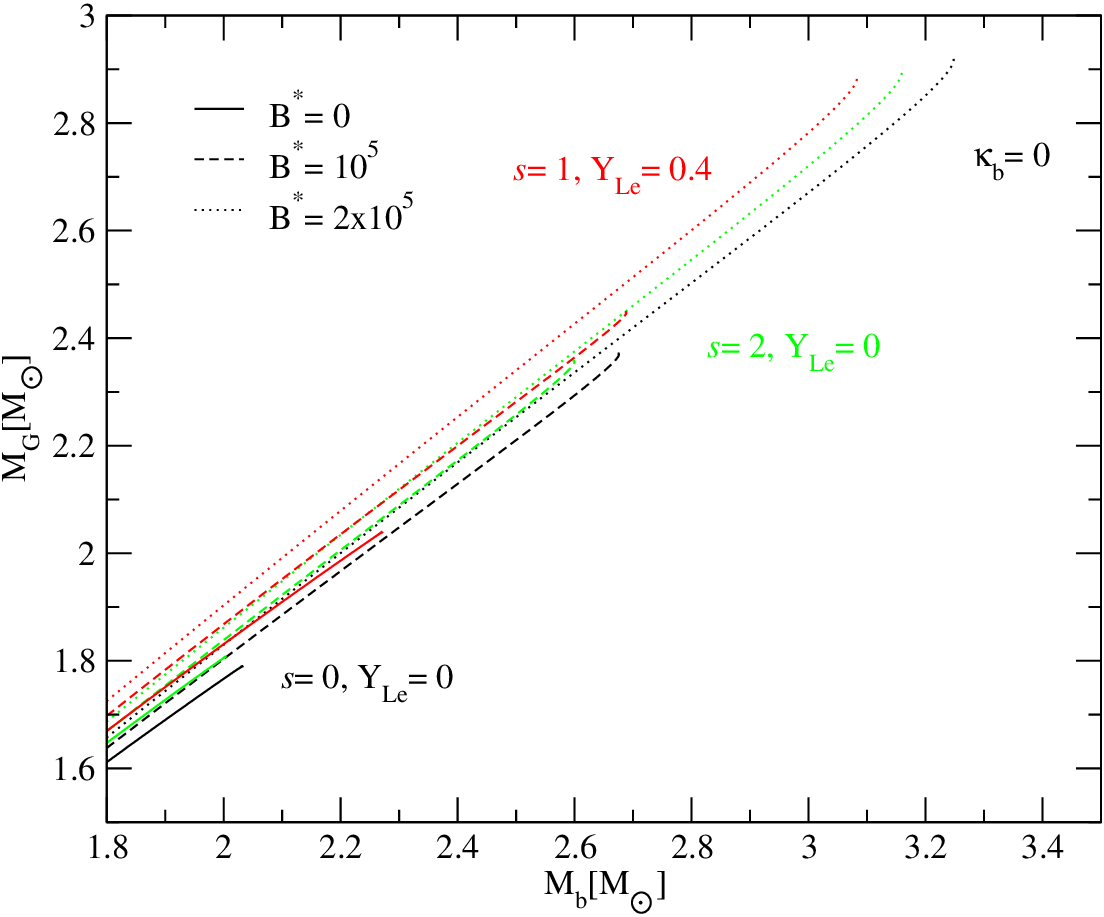}
\end{tabular}
\caption{(Color online) Gravitational mass as a function of the baryonic mass of neutron 
stars for several values of the magnetic
field, using a density dependent magnetic field $B$ given by Eq.~(\ref{brho}). }
\label{massgrav2}
\end{figure}

At zero magnetic field, the maximum baryonic mass of a star with trapped neutrinos and
${\mathfrak{s}}=1$ is 2.27 $M_\odot$. This star will deleptonize and heat up. However, since
the maximum baryonic mass of a neutrino free star with ${\mathfrak{s}}=2$ is  0.27 $M_\odot$
smaller (equal do 2.00 $M_\odot$ the star will evolve into a low-mass black hole
\cite{prakash97}). This is due to the softening that the EOS suffers  with the appearence of
hyperons and  is well illustrated in Fig. \ref{massgrav2} by the full lines: for $B=0$ all
configurations with trapped neutrinos and a baryonic mass above the maximum of the neutrinos
free ${\mathfrak{s}}=2$ configurations, 2.024 $M_\odot$, will evolve into a black hole. 

We will now consider that the decay of the magnetic field will occur in longer timescale than
the deleptonization phase, and therefore, during  the star evolution the
magnetic field remains constant. If we consider $B^*=10^5$, the maximum baryonic mass of a star with
trapped neutrinos and ${\mathfrak{s}}=1$ is 2.69 $M_\odot$. The maximum mass of a neutrino free star with
${\mathfrak{s}}=2$ is  smaller but the difference is much smaller that the one occuring at $B=0$: a
maximum mass of 2.60 $M_\odot$ corresponds only to a 0.09 $M_\odot$ difference. The set of
stars that will evolve to a low mass black hole is much smaller. In  Fig. \ref{massgrav2}, the
dashed curves correspond to $B^*_0=10^5$, from top to bottom ${\mathfrak{s}}=1$ with trapped neutrinos,
neutrino free ${\mathfrak{s}}=2$ and ${\mathfrak{s}}=0$. The configurations of star with trapped neutrinos and ${\mathfrak{s}}=1$
above the maximum of the neutrino free ${\mathfrak{s}}=2$ curve,  2.597 $M_\odot$, will evolve to a
blackhole. However, if we consider $B^*_0=2\times 10^5$ all configurations with trapped neutrinos
and ${\mathfrak{s}}=1$ will evolve to stable ${\mathfrak{s}}=2$ and afterwards ${\mathfrak{s}}=0$ neutrino free star configurations. The maximum mass star configuration with trapped neutrinos and ${\mathfrak{s}}=1$ is smaller than the maximum mass neutrino free star  
 with ${\mathfrak{s}}=2$, respectively 3.07 $M_\odot$ and 3.15 $M_\odot$. No evolution to a low mass blackhole will occur. This could be expected since we have seen discussed that the magnetic field hinders the appearance of hyperons.

However, it is important to notice that if the star cools down in a stable star keeping the magnetic field configuration described by Eq. (\ref{brho}) with $B^*=2\times 10^5$ it may still decay into a low mass blackhole during the magnetic field decay.

\begin{table*}[htb]
\caption{Properties of the stable baryon star with maximum mass,  for several
values of magnetic field using [see  Eq. (\ref{brho})] with the
parametrisation. $M_{max}$, $M^b_{max}$, R, $E_{0}$, $\rho^c$, $B_{c}$,
and $T_c$
are, respectively,  the gravitational and baryonic masses, the star radius,
the central energy density, the central baryonic density, and the values of
the magnetic field  and the temperature at the centre. Neutrino-free matter.}
\label{table3}
\begin{ruledtabular}
\begin{tabular}{ccccccccc}
$B^{*}_{0}$ & ${\mathfrak{s}}$ & $M_{max} [M_\odot]$ & $M^b_{max} [M_\odot]$ &
R [km] &
$E_{0}[\hbox{fm}^{-4}]$ & $\rho_c(\hbox{fm}^{-3})$ &  B$_{c}$(G)  &
$T_c$(MeV)  \\
\hline
$B = 0$                                       &  0 & 1.790 & 2.033 &
11.527 & 5.939 & 0.985 &   -   &  -     \\
                                                    &  1 & 1.794 & 2.024 &
11.717 & 5.854 & 0.967
&   -   &  19.24  \\
                                                    &  2 & 1.808 & 2.004 &
12.467 & 5.656 & 0.922
&   -   &  40.79   \\
$B^{*}_{0} = 10^{5}$             &  0 & 2.372 & 2.672 & 12.694 & 4.846 &
0.688 & 2.812$\times 10^{18}$ & -\\
                                                    &  1 & 2.368 & 2.652 &
12.796 & 4.860 & 0.687
& 2.808$\times
10^{18}$ & 17.96 \\
                                                    &  2 & 2.358 & 2.597 &
13.269 & 4.837 & 0.674
& 2.746$\times
10^{18}$ & 37.43  \\
$B^{*}_{0} = 2\times 10^{5}$&  0 & 2.926 & 3.234 & 14.509 & 3.629 & 0.454
& 3.150$\times 10^{18}$ & - \\
                                                    &  1 & 2.919 & 3.207 &
14.630 & 3.616 & 0.451
& 3.117$\times
10^{18}$ & 15.92 \\
                                                    &  2 & 2.902 & 3.147 &
15.032 & 3.612 & 0.445
& 3.040$\times
10^{18}$ & 33.54 \\
\end{tabular}
\end{ruledtabular}
\end{table*}

\begin{table*}[htb]
\caption{Properties of the stable baryon star with maximum mass,  for several
values of magnetic field using [see  Eq. (\ref{brho})] with the
parametrisation. $M_{max}$, $M^b_{max}$, R, $E_{0}$, $\rho^c$, $B_{c}$,
and $T_c$
are, respectively,  the gravitational and baryonic masses, the star radius,
the central energy density, the central baryonic density, and the values of
the magnetic field  and the temperature at the centre. Neutrino-trapped
matter.}
\label{table4}
\begin{ruledtabular}
\begin{tabular}{ccccccccc}
$B^{*}_{0}$ & ${\mathfrak{s}}$ & $M_{max} [M_\odot]$ & $M^b_{max} [M_\odot]$ &
R [km] &
$E_{0}[\hbox{fm}^{-4}]$ & $\rho_c(\hbox{fm}^{-3})$ &  B$_{c}$(G)  &
$T_c$(MeV)  \\
\hline
$B = 0$                                       &  0 & 2.046 & 2.293 &
12.455 & 5.420 & 0.856 &   -   &  -     \\
                                                    &  1 & 2.040 & 2.271 &
12.529 & 5.395 & 0.847
&   -   &  16.21   \\
                                                    &  2 & 2.036 & 2.226 &
13.198 & 5.192 & 0.808
&   -   &  34.78 \\
$B^{*}_{0} = 10^{5}$             &  0 & 2.454 & 2.694 & 13.123 & 4.819 &
0.660 & 2.678$\times 10^{18}$ & -\\
                                                    &  1 & 2.449 & 2.686 &
13.122 & 4.842 & 0.662
& 2.682$\times
10^{18}$ &  16.76 \\
                                                    &  2 & 2.441 & 2.641 &
13.622 & 4.731 & 0.644
& 2.593$\times
10^{18}$ &  34.72 \\
$B^{*}_{0} = 2\times 10^{5}$&  0 & 2.889 & 3.063 & 14.547 & 3.886 & 0.464
& 3.261$\times 10^{18}$ & - \\
                                                    &  1 & 2.890 & 3.071 &
14.621 & 3.834 & 0.461
& 3.228$\times
10^{18}$ & 16.81 \\
                                                    &  2 & 2.897 & 3.058 &
15.094 & 3.720 & 0.451
& 3.117$\times
10^{18}$ & 33.83 \\
\end{tabular}
\end{ruledtabular}
\end{table*}

\section{Conclusions}
In the present work, we have studied the effect of a very  strong magnetic field on the
EOS and properties of warm  stars. We have  used a relativistic mean 
field model with the GM1 parameter set \cite{gm91} and considered stellar matter both neutrino free and with trapped neutrinos.

Previously, it was shown that there is a strong neutrino suppression at low densities for
finite magnetic fields \cite{aziz10}. For a finite entropy also at high
densities a strong magnetic field gives rise to a decrease of the neutrino fraction,  due to the
larger proton fraction that favors a larger electron fraction. A smaller neutrino fraction may
imply a slower cooling of the star core, since the core cools essentially by neutrino emission
\cite{prakash97}. Another important effect of the magnetic field for a finite entropy per
baryon is the faster reduction of the effective mass with density even when AMM is taken into account.
For zero and finite temperature the effect of $B$ in neutrino free matter is mainly a reduction of
the strangeness with the increase of $B$. However, in matter with  trapped neutrinos, the opposite may occur and  for ${\mathfrak{s}}=2$  the larger the magnetic field  the larger is the strangeness fraction. In neutrino free matter the effect of temperature is not strong enough to oppose the shift of the onset of strangeness to larger densities due to the magnetic field.

It has been shown that a strong magnetic field 
increases the mass and radius of the most massive cold stable star
configuration \cite{broderick1}.  This is still true for warm stars. The radius of these stars
increases with $\mathfrak{s}$, just as it occurs for $B=0$ \cite{prakash97,menezes03}, but in a
much smaller rate. On the other side,  their baryonic masses decrease with the entropy per particle, and the larger decrease occurs for the larger magnetic field. For stronger magnetic fields the contribution of the magnetic field to the total EOS is larger giving rise to a stiffer EOS. As a result,  the star central energy density and baryon density decreases as the magnetic field increases.

The mass of the observed neutron stars may set an upper limit
on the possible magnetic field acceptable in the interior of a star. Of course it may also
occur that the most massive stars decay into low mass blackholes when the magnetic field in
there interior decays. For a lower value of the magnetic field, the EOS becomes softer due to
the onset of hyperons and a smaller maximum baryonic mass results. 

For $B=0$ an hybrid star may evolve to a low mass blackhole because the maximum baryonic mass of a warm star with trapped neutrinos and an entropy per particle ${\mathfrak{s}}\sim 1$ 
is larger than the maximum mass of a warm deleptonized star with ${\mathfrak{s}}=2$ or a cold  deleptonized star \cite {prakash97,menezes04}. In the present study, it was shown that for a strong enough magnetic field, the star would cool down as a stable compact star, if the magnetic field does not decay during the deleptonization phase. However, the decay of the magnetic field may cause star instability and, consequently, the formation of a blackhole.

\begin{acknowledgments}
This work was partially supported by FCT/FEDER under  Projects PTDC/FIS/113292/2009 and CERN/FP/116366/2010,
 and  by COMPSTAR, an ESF Research Networking Programme.
\end{acknowledgments}

\end{document}